\documentclass[fdp,a4paper,fleqn]{w-art}
\usepackage{times}
\usepackage{w-thm}
\usepackage{amssymb}
\usepackage{amsmath}
\usepackage{amsfonts}
\usepackage{array}

\font\helv=cmssbx10

\def\beeta{\pmb{\eta}}
\def\bzeta{\pmb{\zeta}}

\def\mh{\mathcal{H}}

\newcommand{\bk}[2]{\langle #1|#2\rangle}

\usepackage[]{graphicx}
\begin{document}
\DOIsuffix{theDOIsuffix}
\Volume{}
\Issue{}
\Month{}
\Year{}
\pagespan{1}{}
\keywords{Random matrices, Cyclic matrices, Pseudo-Hermiticity, Random walk}



\title[pH-RMT]{Pseudo-Hermitian Random matrix theory}


\author[SCL Srivastava]{Shashi C. L. Srivastava\inst{1,}%
  \footnote{ E-mail:~\textsf{shashi@vecc.gov.in}, 
            Phone: +91\,33\,2318\,4463}}
\address[\inst{1}]{RIBFG, Variable Energy Cyclotron Centre, 1/AF Bidhan nagar, Kolkata-700 064}
\author[SR Jain]{Sudhir R. Jain\inst{2,}\footnote{Corresponding author\quad  E-mail:~\textsf{srjain@barc.gov.in}, 
            Phone: +91\,22\,2559\,3589}}
\address[\inst{2}]{NPD, Bhabha Atomic Research Centre, Mumbai-400 085}
\begin{abstract}
Complex extension of quantum mechanics and the discovery of
pseudo-unitarily invariant random matrix theory has set the stage for a
number of applications of these concepts in physics. We briefly review the
basic ideas and present applications to problems in statistical mechanics
where new results have become possible. We have found it important to
mention the precise directions where advances could be made if further
results become available.
\end{abstract}
\maketitle                   





\section{Introduction}

The postulate of quantum mechanics which makes a contact with measurements is the existence of real eigenvalues of a linear operator corresponding to a physical observable \cite{dirac}. A complex extension of quantum mechanics has been realized where certain special non-Hermitian operators are shown to possess real eigenvalues \cite{caliceti,bender,am,ali2010}. Fundamental concepts like uncertainty relations in quantum mechanics \cite{srjainq} and Poincar\'{e}-Cartan integral invariant \cite{srjainc} have been generalized. A number of examples and formal results have been obtained \cite{tateo}, matrix representations, pseudo-unitary symmetry and group structure has also been established \cite{aj}. On this basis, random matrix theory was presented \cite{aj,jain}. We are thus given a powerful combination of ``non-hermiticity" and ``random matrices"  to model and describe various physical phenomena. The relevance of non-hermiticity has been well-known in many-body theory \cite{rotter}, non-equilibrium 
statistical mechanics \cite{stinchcombe,kirone}, open quantum systems \cite{sieber}, and number theory \cite{takloo}. 

Non-hermiticity of an operator is usually accompanied by a physical interpretation in terms of dissipation. At a rigorous level, the breakdown of Hermiticity naturally occurs when a part of the system is not taken into account, as it corresponds to certain irrelevant variables. Mathematically, this was fortified as the Feshbach projection operator technique. This method has been very successfully employed, particularly in nuclear physics \cite{rotter}. This development has led to an understanding of many aspects of nuclear structure and reactions via what is known as ``shell model embedded in continuum". The relevance of this is over the entire many body theory. To understand this, we just need to recall the first nontrivial instance where a localized state is coupled to a non-interacting fermionic system via one-body operators. The answer found there goes on to build our understanding of the Anderson model of a magnetic impurity residing in a metallic host \cite{bruus}.  

From the first treatment of Brownian motion \cite{einstein}, random walks play a very important role in topics as diverse as polymer physics \cite{degennes}, to locomotion of bacteria \cite{berg,ecoli}. Systems out of equilibrium are described by a master equation with a non-hermitian operator, exemplified by the transfer matrix of the random walk problem. We will review the role played by the recent advances in RMT to this subject.  

RMT is also intimately connected to exactly solvable models, recognized by Sutherland \cite{sutherland}. The generalization of the projection method employed to establish this connection was accomplished and new model was presented \cite{jain2006} for the pseudo-Hermitian case. Of great importance is the random Ising model - we shall argue here that advances in RMT are needed to solve this problem. 

\section{RMT for $2\times2$ pseudo-Hermitian matrices}
In any physical experiment what we measure is a (real) eigenvalue of some operator belonging to a normed vector space. Let us consider a linear wave theory (quantum mechanics) where a system is represented at any instant by a state in a linear normed vector space, with a time-independent norm. Denoting by $|\psi \rangle $ a vector and its conjugate partner by $\langle \psi \eta |$, the inner product is given by the volume integral, $\int \psi ^*\eta \psi dV = \mbox{constant}$. Accordingly, the linearity of evolution and a normed space of functions implies
\begin{eqnarray}
\nonumber
\frac{d}{dt}\int \psi^*\eta\psi dV &=& 0,\\\nonumber
\int \frac{\partial \psi^*}{\partial t}\eta \psi dV &+& \int \psi^*\eta \frac{\partial \psi}{\partial t}dV=0, \\\nonumber
\int i\psi^* \mh^\dag \eta \psi dV &-& \int i\psi^*\eta  \mh \psi dV =0, \\\nonumber
\int i\psi^* \eta (\eta^{-1}\mh^\dag \eta - \mh) \psi dV &=&0, \\ 
\mh^\dag &=&\eta \mh \eta^{-1}.
\end{eqnarray}
We have made use of linearity (via, e.g., Schr\"{o}dinger equation)
\begin{equation}
\mh \psi = i \hbar \frac{\partial \psi}{\partial t}. 
\end{equation}

Let us consider symmetry transformations which preserve the $\beeta$-norm $\left(\bk{x}{\beeta y}\right)$ between the vectors ${\bf x}$ and ${\bf y}$. By considering the Cayley form, {\helv D} = $e^{i\mbox{\helv H}}$ as a symmetry transformation acting on {\bf x, y} where {\helv H} is pseudo-Hermitian in accordance with 
$\beeta${\helv H}$\beeta ^{-1}$ = {\helv H}$^{\dagger}$, it is easy to show that {\helv D} will be pseudo-unitary.  Also it preserves the $\beeta$-norm and a consistently-defined matrix element provided $\mbox{{\helv D} {\bf\helv A} {\helv D}}^{-1}=\mbox{{\helv A'}}$, and is indeed a symmetry transformation \cite{aj}. In general, metric for {\helv H} and {\helv D} are different and we will denote metric for {\helv H} by $\beeta$ and for {\helv D} by $\bzeta$. The metric will always be Hermitian but not unique. Additionally if metric is positive definite, then so is  pseudo-norm. In contrast, any other metric leads to an indefinite pseudo-norm. The closure law for two pseudo-unitary matrices are trivial if they both are pseudo-unitary with respect to  the same metric {\it i.e.} $\beeta = \bzeta$. Also, {\helv D}$^{-1}$ is pseudo-unitary with respect to $\beeta$ if {\helv D} is pseudo-unitary: $\beeta ^{-1}(e^{-i\mbox{\helv H}})^{\dagger}\beeta$ = 
$e^{i\beeta ^{-1}\mbox{\helv H}^{\dagger}\beeta}$ = $e^{i\mbox{\helv H}}$. With identity matrix as unit element of the symmetry transformation, and associativity guaranteed, the N $\times$ N pseudo-unitary matrices form a pseudo-unitary group of order N, $PU(N)$. 

A pseudo-Hermitian operator possesses either real eigenvalues or the eigenvalues occur as complex-conjugate pairs. While this is true in general, the essential point is regarding the corresponding eigenfunctions. As discussed above, the right eigenfunctions have a corresponding left-partner which is related via $\beeta $ and conjugation. Precisely, for ${\cal PT}$-symmetric system, complex conjugation corresponds to time-reversal and parity corresponds to $\beeta $. Connections of 2 $\times $ 2 matrices with ${\cal PT}$-symmetry has been known \cite{aj,jalonso,wang}. The eigenfunctions corresponding to the real (complex conjugate) eigenvalues are (not) simultaneously eigenfunctions of ${\cal PT}$. The eigenfunctions corresponding to the complex conjugate eigenvalues have a zero pseudo-norm \cite{ali2010}. Whereas the real eigenvalues correspond to bound states, the complex conjugate pairs correspond to exceptional points. The latter are very interesting objects which have found way into our understanding of 
quantum phase transitions \cite{cejnar}, zero-width resonances \cite{muller}, quantum chaos \cite{heiss}, multichannel resonance ionization \cite{lefebvre} etc.       

To understand the deviation from Hermiticity while still possessing real eigenvalues, let us take the example of $2\times 2$ matrices where the eigenvalues can be written as
\begin{equation}\label{eq:2deig}
E_{\pm} = \frac{1}{2} \mbox{Tr}H \pm \frac{1}{2} \sqrt{(\mbox{Tr}H)^2-4\mbox{Det}H}.
\end{equation}

Now consider the example matrix,
\begin{equation}
\left[\begin{array}{cc}a&-i \epsilon c\\ic/\epsilon& b\end{array}\right]. 
\end{equation}
It is clearly not Hermitian as off-diagonal elements of the matrix are written as differently scaled elements of corresponding Hermitian matrix with $\epsilon=1$. Now as we have not changed the trace and determinant of the matrix, from Eq. \ref{eq:2deig}, the eigenvalues will be same to that of $\epsilon =1$ (Hermitian) case and so they are real despite the matrix being non-Hermitian. 

Starting with the simplest case of a pseudo-Hermitian matrix \cite{jain},  
\begin{eqnarray}
\mbox{\helv H} = \{\mbox{\helv H}_{ij}\}&= \left[\begin{array}{cc}a&-ib\\ic&a\end{array}\right],
\end{eqnarray}      
$a, b, c$ being real. It is easy to show that metric is   
\begin{eqnarray}
\beeta&=\left[\begin{array}{cc}0&i\\-i&0\end{array}\right]. 
\end{eqnarray}  
Interpreting this metric as parity (${\cal P}$), and usual complex conjugation, ${\cal K}_0$ as time-reversal operator ${\cal T}$, the matrix is ${\cal PT}$- symmetric.  The diagonalizing matrix for {\helv H} is given by 
{\helv D}, {\it i.e.}, 
\begin{equation} 
\mbox{\helv D} =\left[\begin{array}{cc}1&i/r\\ir&1
\end{array}\right].
\end{equation}
The eigenvalues of {\helv H} are $E_\pm=a \pm \left[\frac{c}{2r} + \frac{br}{2} \right]$ ($r = \sqrt{c/b}$ ($r \in [0, \infty ]$ provided $bc>0$)).  Here, as we have  mentioned earlier  that {\helv H}({\helv D}) may not be pseudo-Hermitian (unitary) with respect to same metric, the metric for {\helv D} is different from to that of {\helv H} and is given by 
\begin{eqnarray}
\bzeta&=\left[\begin{array}{cc}0&1\\1&0\end{array}\right]. 
\end{eqnarray} 

In the parameter space, the joint probability density function for the  matrix {\helv H} \cite{mehta}
\begin{equation}
P(\mbox{\helv H}) = {\cal N} e^{- \frac{1}{2\sigma ^2}~tr~\mbox{\helv H}^{\dagger}
\mbox{\helv H}}
\end{equation}
is taken of the Wishart form, and this reduces to

\begin{equation}    
P(a, b, c) = {\frac{1}{2 (\pi \sigma ^2)^{\frac{3}{2}}}} e^{-\frac{1}{2\sigma ^2} 
\left[2a^2 + b^2 + c^2 \right]}. 
\end{equation}               
By inserting  the relation of eigenvalues in terms of parameters and, using the Jacobian for this transformation, we get  the joint probability distribution function (jpdf) of eigenvalues:  
\begin{equation}    
P(E_+, E_-) = \frac{|E_+ - E_-|}{2 (\pi \sigma ^2)^{\frac{3}{2}}} 
K_0\left(\frac{(E_+ - E_-)^2}{4\sigma ^2}\right) 
e^{-\frac{(E_+ + E_-)^2}{4\sigma ^2}}. 
\end{equation}

Perhaps historically one of the most studied quantity in random matrix literature is the nearest neighbour level spacing distribution, $P(S)$. It is well known that for the Wigner-Dyson ensembles the spacing distribution very well-approximated by, $P(S) \sim S^{\beta}e^{-\gamma S^2}$ where $\beta$ is 1, 2, and 4 corresponds  to the orthogonal, unitary, and symplectic ensembles \cite{mehta,haake,zelevinsky}. However, there are systems such as billiards in polygonal enclosures, three-dimensional Anderson model at the metal-insulator transition point, 
and many more those display intermediate statistics \cite{parab,gremaud,bgs}. 

The spacing distribution for the present case , $P(S)$, is given in terms of the jpdf by 
\begin{eqnarray}
P(S) &=& \int_{-\infty}^{\infty}\int_{-\infty}^{\infty}P(E_+, E_-)
\delta (S - |E_+ - E_-|)dE_+dE_- \nonumber \\
&=& \frac{|S|}{\pi\sigma ^2}K_0\left(\frac{S^2}{4\sigma ^2}\right).
\end{eqnarray}  
Remarkably, as $S \to 0$,  $P(S) \sim S \log (1/S)$ (it is non-algebraic level repulsion, in contrast with Wigner-Dyson ensembles). In Table.\ref{tab:2by2}, we summarize the results for various $2\times 2$ pseudo-Hermitian matrices which form their own class. 

\begin{table}[ht]
\caption{A summary of $2 \times 2$ pseudo-hermitian random matrices} 
\centering  
\begin{tabular}{c c m{1.5in} c c} 
\hline\hline                        
{\helv H} & $\beeta$  & \begin{center}{\helv D}\end{center} & $\bzeta$ & P(S) ($S \to 0$)  \\ [0.5ex] 
\hline                
$\begin{bmatrix}
a & -i b\\
ic & a
\end{bmatrix}$  & 
$\begin{bmatrix}
0 & i \\
-i & 0
\end{bmatrix}$ & \begin{center}
$\frac{1}{\sqrt{2}}\begin{bmatrix}
1 & \frac{i}{r}\\
ir & 1
\end{bmatrix}$ $(0 \le r <\infty)$ \end{center} & 
$\begin{bmatrix}
0 & 1\\
1 & 0
\end{bmatrix}$ & 
$-S\log S$ \\

$\begin{bmatrix}
a+c & ib\\
ib & a-c
\end{bmatrix}$  & 
$\begin{bmatrix}
1 & 0 \\
0 & -1
\end{bmatrix}$ & \begin{center}
$\frac{1}{\sqrt{\cos 2\theta}}\begin{bmatrix}
\cos \theta & i \sin \theta\\
-i \sin \theta & \cos \theta
\end{bmatrix}$ $(-\pi/4 < \theta <\pi/4)$ \end{center}& 
Not known & 
$-S\log S$ \\

$\begin{bmatrix}
a & -i \epsilon c\\
\frac{ic}{\epsilon} & b
\end{bmatrix}$  & 
$\begin{bmatrix}
\frac{1}{\epsilon} & 0 \\
0 & \epsilon
\end{bmatrix}$ & \begin{center}
$\begin{bmatrix}
\cos \theta & i \epsilon\sin \theta\\
-i \sin \theta/\epsilon & \cos \theta
\end{bmatrix} $ \end{center}& 
$\begin{bmatrix}
\frac{1}{\epsilon} & 0 \\
0 & \epsilon
\end{bmatrix}$ & 
$Sf(\gamma), \epsilon = e^\gamma$ \\

$\begin{bmatrix}
a+ib & c\\
d & a-ib
\end{bmatrix}$  & 
$\begin{bmatrix}
0 & 1\\
1 & 0
\end{bmatrix}$ & \begin{center}
$\begin{bmatrix}
\frac{r e^{i\theta}}{\sin\theta} & -\frac{r e^{i\theta}}{\sin\theta}\\
1& 1
\end{bmatrix}$ \end{center}& 
Not known & 
$S$ with large slope \\

$\begin{bmatrix}
a+b & d+ic\\
-d +ic & a-b
\end{bmatrix}$  & 
$\begin{bmatrix}
1 & 0\\
0 & -1
\end{bmatrix}$ & \begin{center}
$\begin{bmatrix}
i \cos\theta &  e^{i\theta}\sin\theta\\
 e^{-i\theta}\sin\theta & -i \cos\theta
\end{bmatrix}$ \end{center}& 
$\begin{bmatrix}
1 & 0\\
0 & -1
\end{bmatrix}$ & 
$S$  
\\[1ex]      
\hline 
\end{tabular}
\label{tab:2by2} 
\end{table}

It is clear that unlike Hermitian cases, simply because of the presence of a larger number of parameters, a wide spectrum of spacing distributions are expected. The natural extension of these results to the general $PU(N)$ case remains open. However, we  have  exact results for a special kind of pseudo-Hermitian matrices, {\it viz.} cyclic matrices or circulants. 

\section{Cyclic Matrices and random matrix theory}

Let us consider an $N \times N$ cyclic matrix with real elements, $\{a_i\}$:
\begin{eqnarray}\label{eq:1}
\mbox{\helv M} = \left[\begin{array}{cccc}a_1&a_2&...&a_N\\a_N&a_1&...&a_{N-1}\\ \vdots & & & \\a_2&a_3&...&a_1\end{array}\right].
\end{eqnarray}   
It is important to note that this matrix is, in fact, pseudo-Hermitian (pseudo-orthogonal) with respect to $\beeta $ \cite{RCM}
\begin{eqnarray}\label{eq:2}
\beeta = \left[\begin{array}{cccccc}1&0&0&...&0&0\\0&0&0&...&0&1\\0&0&0&...&1&0 \\\vdots & & & & & \\0&1&0&...&0&0\end{array}\right],
\end{eqnarray}   
that is, 
\begin{equation}\label{eq:3}
\mbox{\helv M}^{\dagger} = \mbox{\helv M}^T = \beeta \mbox{\helv M} \beeta ^{-1}.
\end{equation}
Since $\beeta ^2 = $ identity, {\helv I}, consistent with the earlier discussion, $\beeta $ may be called ``generalized parity". We can distribute the matrix elements using (7), $P({\bf M} \sim \exp [-A Tr ({\bf M}^{\dagger}{\bf M})])$, and obtain an 
ensemble of random cyclic matrices (RCM) that is pseudo-orthogonally invariant.     It is a well-known \cite{kowaleski} that a cyclic matrix is  diagonalized by a Fourier matrix, ${\bf U}$, which is unitary:

\begin{equation}\label{eq:fourier}
U_{jl} = \frac{1}{\sqrt{N}} \exp \frac{2\pi i}{N}(j-1)(l-1).
\end{equation} 
The eigenvalues of {\helv M} are given by \cite{kowaleski}
\begin{equation}\label{eq:4}
E_{l} = \sum_{p=1}^{N} a_p \exp \left[\frac{2\pi i}{N}(p-1)(l-1)\right]; 
\end{equation} 
($l = 1, 2, ..., N$), there is always one real eigenvalue as a sum of the elements, due to permutation symmetry.

Let's start with the analysis for an ensemble of $3 \times 3$ cyclic matrices. To derive the  joint probability distribution function (JPDF)of eigenvalues, as all the correlations are related to it we again need to calculate the Jacobian of transformation from parameters of matrices to the eigenvalues.
We immediately see that ${\mathrm tr~} \mbox{\helv M}^{\dagger} \mbox{\helv M} = 3 (a_1^2 + a_2^2 + a_3^2)$. The JPDF of eigenvalues $P(\{E_i\})$ can be written as 
\begin{eqnarray}\label{eq:jpdf3cyc}
P(E_1, E_2, E_2^*) = \left(\frac{A}{\pi} \right)^{\frac{3}{2}}e^{-A (E_1^2 + 2 |E_2|^2)}.
\end{eqnarray}
As there are also two complex eigenvalues present, the definition of spacing is taken in the sense of Euclidean distance between the eigenvalues. With this definition of spacing, we may define $S_{23} := |E_2 - E_3| = \sqrt{3}(a_3 - a_2)$ as well as $S_{12} := |E_1 - E_2| = |\frac{3}{2}(a_2 + a_3) + \frac{i\sqrt{3}}{2}(a_2 - a_3)|$. Obviously, $S_{12} = S_{13}$ as $E_2$ and $E_3$ are complex conjugate. Spacing distribution for the complex conjugate pair, $P_{cc}(S_{23})$ is given by
\begin{eqnarray}\label{eq:spac_cc}
P_{cc}(S_{23}) &=& \int \prod_{i=1}^{3}da_i P(\{a_i\})\delta (S_{23} - \sqrt{3}|a_3 - a_2|) \\ \nonumber
&=& \sqrt{\frac{2A}{\pi}} e^{-\frac{A}{2}S_{23}^2}.
\end{eqnarray}   
The normalized spacing distribution by setting average spacing as 1, can be written in terms of the variable $z$:
\begin{equation}\label{eq:spac_ccz}
p_{cc}(z) = \frac{2}{\pi} e^{-\frac{z^2}{\pi }}.
\end{equation} 

Similarly, the spacing distribution, $P_{rc}(S_{12})$ in normalized form can be obtained gain in in the variable $z$:
\begin{eqnarray}\label{eq:spac_rcz}
p_{rc}(z)&=&\frac{3\sqrt{3}\pi }{16} c^2z\exp \left(-\frac{3\pi }{16} c^2z^2\right) I_0\left(\frac{3\pi }{32} c^2z^2\right)
\end{eqnarray} 
where $c=~_2F_1[\frac{3}{4},\frac{5}{4},1,\frac{1}{4}]$.

\begin{figure}[h]
\begin{center}
\includegraphics{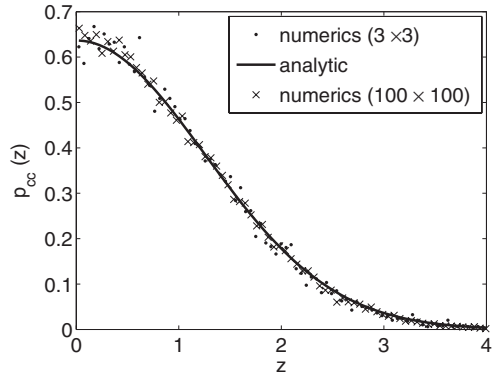}
\end{center}
\caption{Probability distribution of the absolute spacing between the complex conjugate pair of eigenvalues of a 
Gaussian ensemble of $3 \times 3$ cyclic matrices. The numerical result matches well with the analytic result (\ref{eq:spac_ccz}).\label{fig:pcc3by3}}
\end{figure}

\begin{figure}[h]
\begin{center}
\includegraphics[scale=0.6]{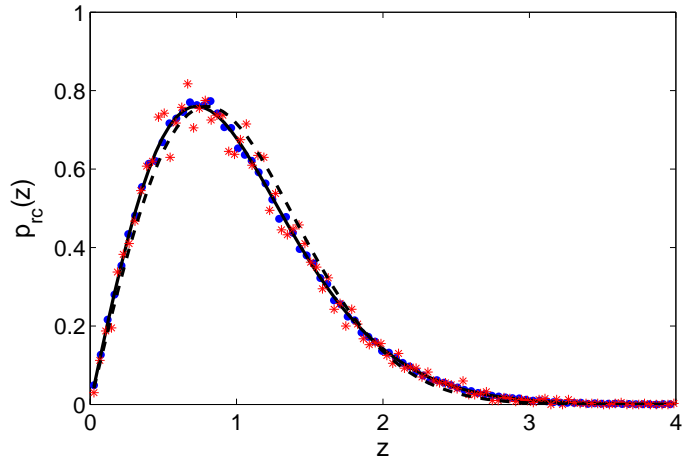}
\end{center}
\caption{Probability distribution of the absolute spacing between a real and a complex eigenvalue of a 
Gaussian ensemble of $3 \times 3$ cyclic matrices (red stars). The numerical result obtained by considering 10000 realizations for $3 \times 3$ matrices and 1000 realizations of $100 \times 100$ matrices (blue circles) agrees with the analytic result (solid black)(Eq. \ref{eq:spac_rcz}). A difference from Wigner distribution for GOE (black dotted) is clearly shown in the figure. \label{fig:prc3by3}}
\end{figure}

It is clear that the Gaussianity of $p_{cc}(z)$ implies that there is neither a level repulsion nor any attraction among the complex conjugate pairs while real and complex eigenvalues display linear level repulsion. The numerical simulations are also in agreement as can be seen in Fig. \ref{fig:pcc3by3} and Fig. \ref{fig:prc3by3}. 

In the same manner the jpdf for the general case of $N \times N$ matrices (with even $N$) is found to be
\begin{eqnarray}\label{eq:jpdfevenN}
P(\{E_i\}) &=& \left( \frac{A}{\pi } \right)^{\frac{N}{2}} \exp \bigg[-A \bigg(E_1^2 + E_{\frac{N}{2}+1}^2 + \sum_{i \neq 1, \frac{N}{2}+1}^{N} E_iE_{N+2-i} \bigg)\bigg]
\end{eqnarray}   
where $E_1$ and $E_{\frac{N}{2}+1}$ real and the rest of the eigenvalues may be complex. For odd $N$, there will be only one real eigenvalue, $E_1$ and the summation in the second term will extend over all $i$ except 1.

\begin{figure}[h]
\begin{center}
\includegraphics{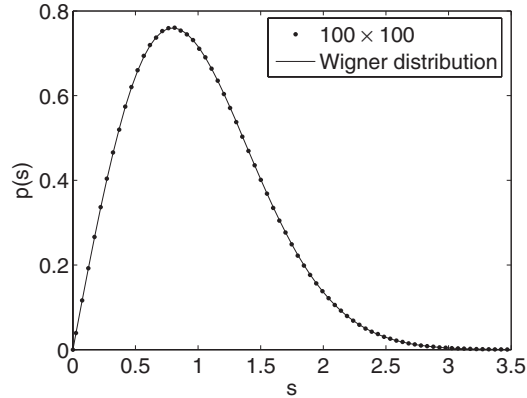}
\end{center}
\caption{ Probability distribution for spacing between two complex eigenvalues which are not complex conjugate. The agreement with analytical result (Eq. \ref{eq:spac_wigner}) is good. \label{fig:prc100by100}}
\end{figure}

For the general case of $N$, one more spacing will appear which was absent in $3\times 3$ case and that is between two complex eigenvalues which are not complex conjugate. Rest are found to be the same as in $3\times 3$ case. This spacing distribution which we denote be simply $p(s)$ in normalized form is found to be
\begin{eqnarray}\label{eq:spac_wigner}
p(s) &=& \frac{\pi s}{2}\exp \left(-\frac{\pi s^2}{4}\right)
\end{eqnarray}
which is exactly the Rayleigh distribution \cite{takloo} (Fig. \ref{fig:prc100by100}).  In contrast to Wigner's result which was an (excellent) approximation for $N$ dimensional symmetric matrices, here it is an exact result for all $N$. Interestingly the same distribution occurs inthe case of a random Poisson point process  in a plane. Due to the fact that Wigner's distribution is approximate for $N \times N$ matrices, the above result is interpreted more gainfully in terms of the Rayleigh distribution of complex eigenvalues.  

As it has been already established that {\helv M} is an example of pseudo-Hermitian (orthogonal) matrix, and as these results are found for $N \times N$ matrices, we believe that the results on cyclic matrices not only extend the random matrix theory in a significant way but also provide an illustrative example for  more general results found for the Ginibre orthogonal ensemble with Gaussian distributed real elements solved recently by \cite{Kanzieper,Peter}. This also serves as an example where degree of freedom for the matrices are very constrained (only $N$), and how that  affects the general resullts.

\section{Biased random walks on a regular lattice and cyclic matrices}

Let us first consider a random walk on a one-dimensional lattice of $N$ equally spaced sites with periodic boundary conditions.  Assuming that the jump probability is $w$, it decides to jump to  left or right neighbour with probabilities $pw$ and $qw (=(1-p)w)$ respectively.  Let us consider an ensemble of such lattices, and define the probability of occupation of the site, $i$ by 
\begin{equation}\label{eq:p_occup}
p_i=N_i /N_{ens}.
\end{equation}  
where $N_i$ denotes the number of lattices (realizations) with a particle occupying the $i^{th}$ site and $N_{ens}=\displaystyle\sum_{i=1}^{N}N_i$. 

At time $t$, a state of an ensemble can be written as a vector
\begin{eqnarray}\label{eq:vecpt}
\vec{p}(t)=\bigg[p_1(t),...  ,p_N(t)\bigg]^T,
\end{eqnarray}
where $T$ stands for transpose.
The time evolution of the ensemble is given by
\begin{eqnarray}\label{eq:vecpt_evol}
\vec{p}(t+1)={\bf M} \vec{p}(t)
\end{eqnarray}
where
\begin{eqnarray}\label{eq:transM}
{\bf M} = \left[\begin{array}{ccccc}(1-w)&pw&0&...&qw\\qw&(1-w)&pw&...&0\\0&qw&(1-w)&...&pw\\...& & & &\\pw&0&qw&...&(1-w)\end{array}\right].
\end{eqnarray}
Matrix ${\bf M}$ is a transition matrix which can be easily recognized as an asymmetric cyclic matrix.  Since this matrix is not Hermitian, its  eigenvalues occur in complex conjugate pairs, in addition to some of them being real. As we know, the Fourier matrix is the  diagonalizing matrix for cyclic matrices. The $n^{th}$ component of the $l^{th}$ eigenvector corresponding to the eigenvalue, $\lambda _l$ is
\begin{eqnarray}\label{eq:ev}
(\hat e_l)_n=\frac{1}{\sqrt{N}}\exp\bigg[\frac{2\pi i}{N}(n-1)(l-1)\bigg].
\end{eqnarray} 
Now, any initial distribution can be expanded in terms of the eigenvectors of cyclic matrix and time evolution simplifies to essentially taking powers of eigenvalues and then recombining which yields
\begin{eqnarray}\label{eq:p_i_t}
p_i(t)=\displaystyle \sum_{l=1}^N c_l \lambda_l^t (\hat e_l)_i
\end{eqnarray} 
This equation is used to calculate the time evolution of $p_i(t)$ for lattices of several different numbers of sites and for given values of $w$.

\begin{figure}[h]
\begin{center}
\includegraphics[width=0.5\linewidth, height=5cm]{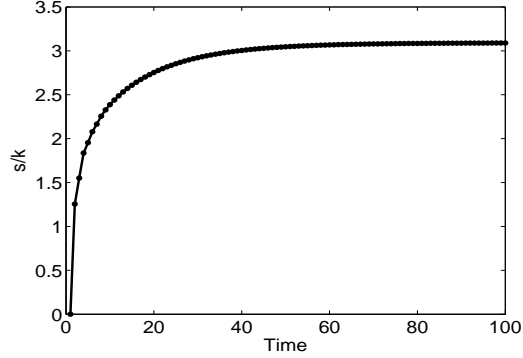}
\end{center}
\caption{For the case of random walkers on a periodic lattice with 22 sites,  $a_1 = 0.2, a_2 = 0.24, a_{22} = 0.56$ (other $a_i$'s are zero) with the  jump probability, $w = 0.8$, the entropy is seen here.}
 \label{srf}
\end{figure}

We will use Boltzmann's relation to calculate the entropy of the system, $S=k_B\ln \Omega $ with the thermodynamic probability, 
\begin{eqnarray}\label{eq:thermProb}
\Omega=\frac{N_{ens}!}{(N_1!)(N_2!)...(N_{N!})}.
\end{eqnarray}
Using Stirling's approximation for large N, and scaling by $N_{ens}$, the ensemble averaged entropy, $s=S/N_{ens}$ is 
\begin{eqnarray}\label{eq:av_ent}
\frac{s}{k_B}=-\displaystyle\sum_{i=1}^{N_{site}}p_i\ln p_i.
\end{eqnarray}
As equilibrium distribution  corresponds to the largest eigenvalue, $p_i=1/N_{site}$, and hence the limiting value of $s/k_B$ is $\ln N_{site}$. The evolution of entropy is shown in Fig.\ref{srf}.   

\section{Biased random walks on a disordered lattice}

We can generalize the previous model by allowing finite jump probability to all the sites, and let us  introduce randomness in the elements of the transition matrix. The question is if transition matrix of the biased
random walk is chosen from a Gaussian ensemble of cyclic matrices, then what can be said about the spectral
properties and hence the  evolution of entropy of this system?

It is important to note that the matrix elements are probabilities and their sum  is unity. Let us
define the average by
\begin{eqnarray}\label{eq:19}
\langle\vec{p}(t)\rangle_{RCM}=\frac{\int_0^{1}dr \int_{-\pi }^{\pi } d\theta \vec{p}(t)\rho (r, \theta ) r }{\int_0^{1}dr \int_{-\pi }^{\pi }d \theta \rho(r, \theta )r}
\end{eqnarray}
where $\rho (r, \theta )$ is the density of eigenvalues $(\lambda = r e^{i\theta})$. It can be shown that  $r$ ($= |\lambda |$) is distributed by a Wigner distribution (Eq. \ref{eq:spac_wigner}) of random matrix theory, by the similar arguments as for Eq. \ref{eq:spac_wigner} (also see\cite{RCM}). 

Using Eq.\ref{eq:p_i_t} and Eq.\ref{eq:ev}, $p_j(t)$ can be re-written as 
\begin{eqnarray}\label{eq:22}
p_j(t)&=&\frac{1}{N}\displaystyle \sum_{n=1}^N  p_n(0)\displaystyle \left (\lambda_1^t+ \sum_{l=2}^N \lambda _l^t\exp \bigg[\frac{2\pi i}{N}(l-1)(j-n)\bigg]\right) \nonumber \\ 
p_j(t)-\frac{1}{N} &=& \frac{1}{N} \sum_{n=1}^N  p_n(0) \sum_{l=2}^{N} \lambda _l^t\Omega _{jn}^{l-1}
\end{eqnarray}
where we have $\exp \bigg[\frac{2\pi i}{N}(j-n)\bigg] := \Omega _{jn}$ and also made use of $\sum_{n=1}^N  p_n(0)=1$. Now, we can average over the the joint probability distribution function of eigenvalues, $J(\lambda _2, ..., \lambda _N )$. Denoting the average 
as $\langle p_j(t)-\frac{1}{N}(\{ \lambda _i \}) \rangle _{RMT}$, we can write

\begin{eqnarray}\label{eq:23}
\langle \tilde{p}_j(t)(\{ \lambda _i \}) \rangle _{RMT} &=& \frac{1}{N} \sum_{n=1}^N  p_n(0) \sum_{l=2}^{N} \Omega _{jn}^{l-1} \int d\lambda _2 ...d\lambda _l ... d\lambda _N J(\lambda _2, ..., \lambda _N )\lambda _l^t \nonumber \\
&=& \frac{1}{N} \sum_{n=1}^N  p_n(0) \sum_{l=2}^{N} \Omega _{jn}^{l-1} \left[ \frac{1}{N-1} \int_{-\infty }^{+\infty } d\lambda  \lambda ^t\rho(\lambda )\right] \nonumber \\ \nonumber
&=& \frac{1}{N (N-1)} \sum_{n=1}^N  p_n(0) \sum_{l=2}^{N} \Omega _{jn}^{l-1}.\left[ \int_{0 }^{+\infty } \int_{-\pi }^{+\pi } k d r d\theta \right.\\ &&\left.  r^t \exp\left[ i t \theta \right] \frac{\pi }{2}r^2 \exp \left( -\frac{\pi r ^2}{4} \right) \left(\frac{1}{\pi}-\delta(\theta)\right)\right] \nonumber\\\nonumber
&=& \frac{1}{N}\frac{1}{\left( -e^{-\pi /4}+\text{Erf}\left[\frac{\sqrt \pi}{2}\right]\right)} \left(\frac{2}{\sqrt\pi}\right)^{1+t}\gamma\left(\frac{3+t}{2},\frac{\pi }{4}\right).
\end{eqnarray}

Using the identity for large $a$ ($\neq 0,-1,-2,...$) and fixed $z$, \cite{dlmf}
\begin{equation}\label{eq:24}
\gamma\left(a,z\right)=z^{a}e^{{-z}}\sum _{{k=0}}^{\infty}\frac{z^{k}\Gamma(a)}{\Gamma(a+k+1)}
\end{equation}
and after a little algebra, we can rewrite  $N \langle \tilde{p}_j(t)(\{ \lambda _i \}) \rangle _{RMT}$ for large time $t$, 
\begin{eqnarray}\label{eq:25}
N \langle \tilde{p}_j(t)(\{ \lambda _i \}) \rangle _{RMT} &\approx & \frac{\pi}{4} \frac{e^{-\frac{\pi}{4}}}{\left( -e^{-\pi /4}+\text{Erf}\left[\frac{\sqrt \pi}{2}\right]\right)} \left[\frac{2}{t+3} +\frac{\pi}{(t+3)(t+5)}+ O\left(\frac{1}{t^3}\right)\right]
\end{eqnarray}

\begin{figure}[h]
\begin{center}
\includegraphics[width=0.5\linewidth, height=5cm]{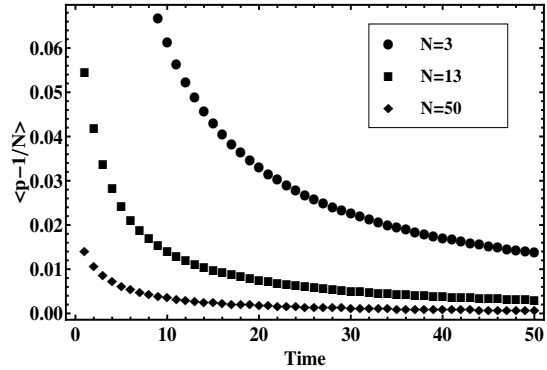}
\end{center}
\caption{With time $p$ becomes more and more uniform, for three different $N$ values the curves show how fast the uniformity sets in with time. }
 \label{rwalk_av}
\end{figure}

\begin{figure}[h]
\begin{center}
\includegraphics[width=0.5\linewidth, height=5cm]{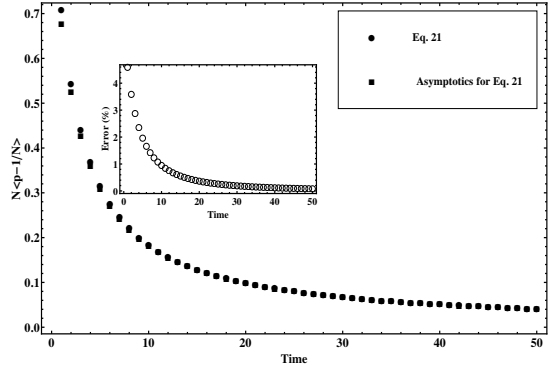}
\end{center}
\caption{After absorbing $N$ in $\langle \tilde{p}_j(t)(\{ \lambda _i \}) \rangle _{RMT}$, the exact time dependence and first two term of asymptotic series expansion is plotted, in inset the difference between the two in $\%$ is plotted. It is clear that the convergence of the series is very fast. Asymptotically, the time dependence is given by $\sim \frac{\rm constant}{t+3}$.}
 \label{rwalk}
\end{figure}

The exclusion of first eigenvalue which always corresponds to $\theta=0$ leads to the exclusion of  the density of this angle ($\delta(\theta))$ from the uniform density of $\theta$, valid for the rest of the eigenvalues.

The time dependence of $\langle \tilde{p}_j(t)(\{ \lambda _i \}) \rangle _{RMT}$ is shown in Fig.\ref{rwalk_av}. 
Approach of all the probabilities to $1/N$ is evident, and so is the approach to a non-equilibrium steady state with maximum entropy given by $\log N$. This is not surprising. Utilizing the connection cyclic matrix models, we believe that not only Gaussian but other kind of randomness can also be modelled. This a clean and simple way to do what could otherwise be done by setting up a master equation with a non-Hermitian Hamiltonian \cite{mallick} and solve for the steady state solution to study the approach to equilibrium.  

\section{Cyclic blocks, Ising Model and pseudo-Hermiticity}

In the two-dimensional Ising model defined on a regular square lattice, we associate a spin variable $\sigma _i$ with values +1 or -1 with each site. The nearest neighbour interaction terms is $J_{ij}\sigma _i\sigma _j$, and zero otherwise. With fixed $J_{ij} = J$, the partition function was found by Onsager \cite{onsager,kaufman}. However, as is well-known (p. 7 of \cite{mehta}), if $J_{ij}$ is a random variable, with a symmetric distribution around zero mean , we have the random Ising model. the calculation of the partition function of which is an open problem. 

In the treatment of Onsager and Kaufman, a very crucial role is played by the transfer matrix with elements containing the coupling which was identified as having cyclic block structure.  It would be nice to study the matrix models with such cyclic block structures. We have already seen that cyclic matrices with scaler entries are pseudo-Hermitian too. Let's see what can be said about cyclic blocks \cite{shashi_blocks}?

In this Section, we present a special case of random matrices with block entries. Firstly, let us consider  
\begin{eqnarray}
\mbox{\helv B} = \left[\begin{array}{cccc}A_1&A_2&...&A_N\\A_N&A_1&...&A_{N-1}\\ \vdots & & & \\A_2&A_3&...&A_1\end{array}\right]
\end{eqnarray}
where each entry $A_i$ is 
\begin{eqnarray}
A_i = \left[\begin{array}{cc}a_i&-b_i\\c_i&a_i\end{array}\right].
\end{eqnarray}
$a_i, b_i,$ and $c_i$  are drawn independently from a Gaussian distribution with zero mean and unit variance. 
The matrix {\helv B} is pseudo-orthogonal with respect to the ``generalized parity", 
\begin{eqnarray}
\Sigma = \left[\begin{array}{cccc}\sigma &0 &...&0 \\0 &0 &...&\sigma \\ \vdots & & & \\0 &\sigma &...&0 \end{array}\right]
\end{eqnarray}
where $\sigma $ is the Pauli matrix,
\begin{eqnarray}
\sigma = \left[\begin{array}{cc}0&1\\1&0\end{array}\right].
\end{eqnarray}
That is, {\helv B}$^{\dagger}=\Sigma ${\helv B}$\Sigma ^{-1}$.

\begin{figure}[h] 
\begin{center}
\includegraphics[scale=0.4]{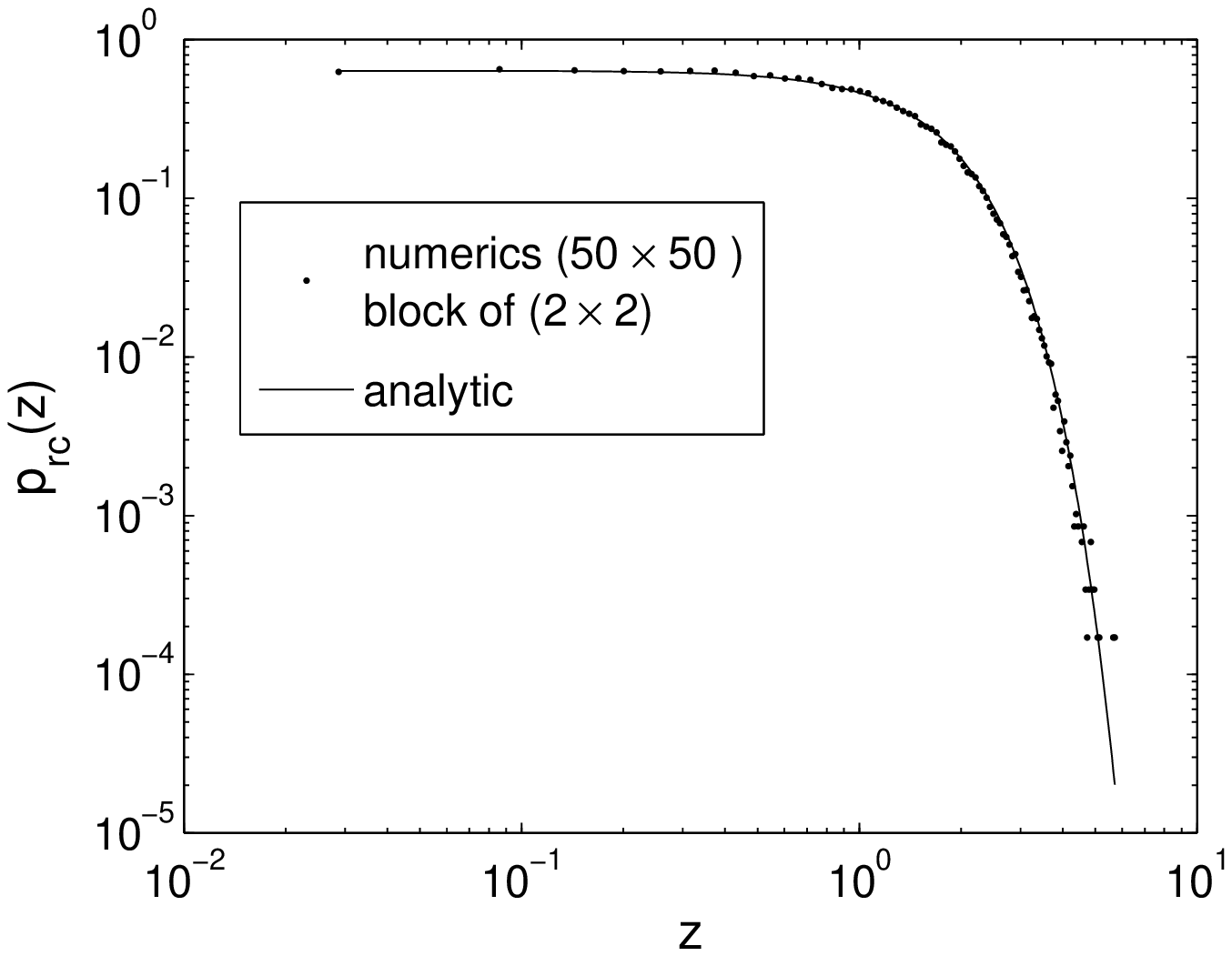}
\includegraphics[scale=0.4]{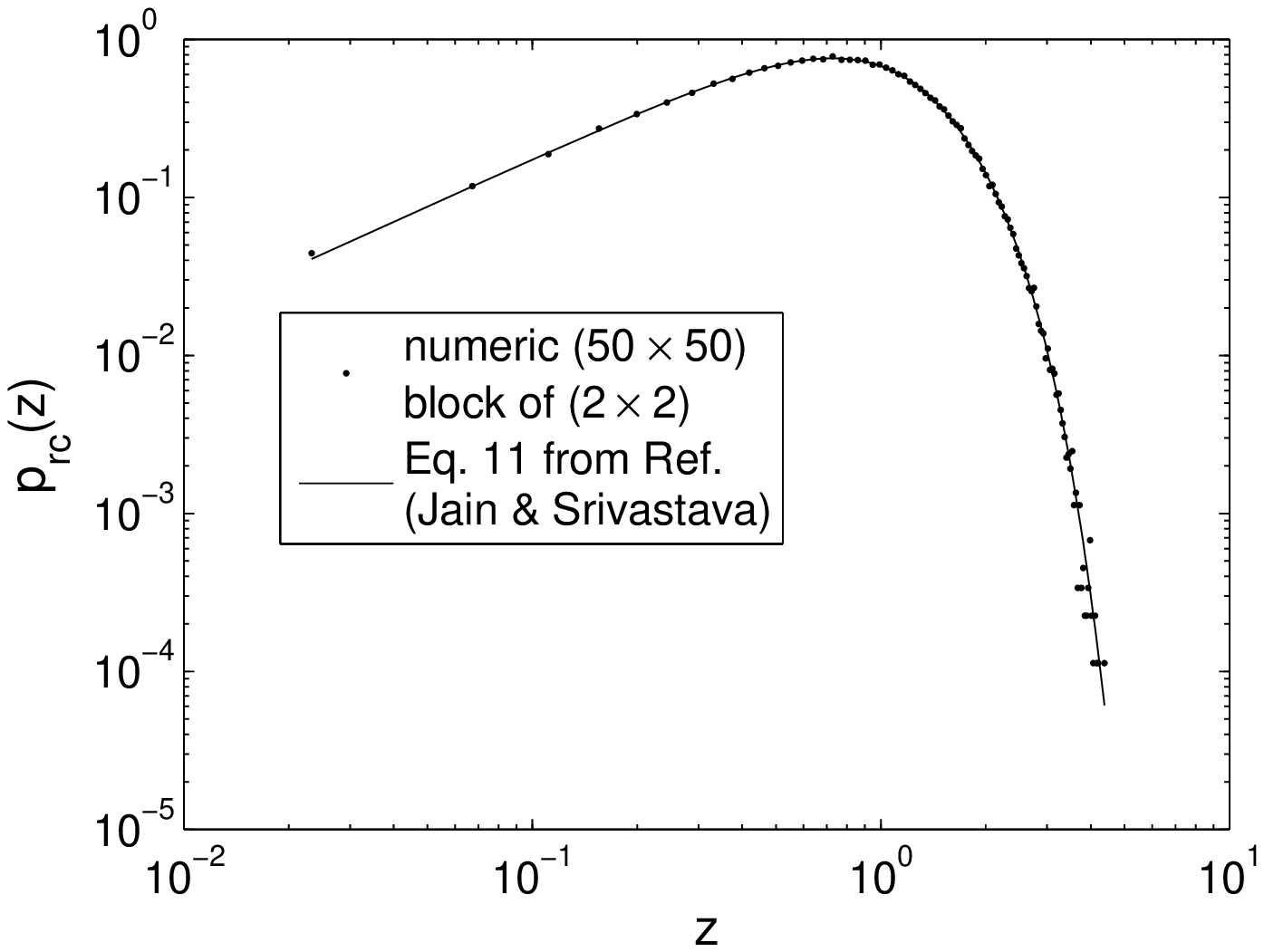}
\end{center}
\caption{(Left panel) Log-log plot of the distribution of spacing among complex-conjugate pairs. The result is exactly in agreement with Eq. \ref{eq:spac_ccz} for scalar entries.(Right panel) Log-log plot of the distribution of spacing among real and complex eigenvalues. The result is exactly in agreement with Eq. \ref{eq:spac_rcz} (Eq 11 from Ref. Jain \& Srivastava) for scalar entries.} \label{gauss_block_asym_ph}
\end{figure}

\begin{figure}[h]
\begin{center}
\includegraphics[scale=0.5]{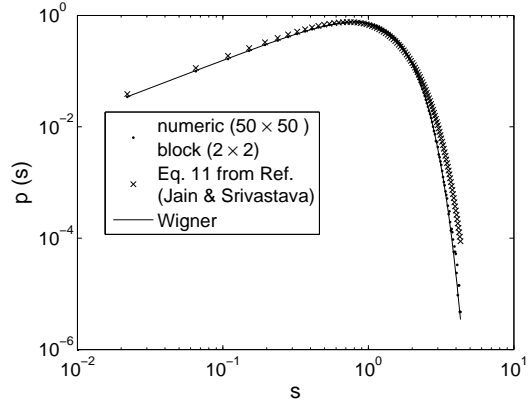}
\end{center}
\caption{Log-log plot of the distribution of spacing among complex eigenvalues. The result is exactly in agreement with Eq. \ref{eq:spac_wigner} for scalar entries (Eq. \ref{eq:spac_rcz} is Eq. 11 from Ref. Jain \& Srivastava).}
\label{wigner_asym_ph}
\end{figure}

However, numerically it is seen that the spectral fluctuations are the same as that for random cyclic matrices with scaler entries. For 50 $\times $ 50 matrix, comprised of fifty 2$\times $2 blocks per row similar spacing distributions among complex conjugate pairs, real-complex pair, complex-complex pair is found to have in good agreement with same in case of cyclic matrices with scaler entry {\it i.e.} Eq.  \ref{eq:spac_ccz}, \ref{eq:spac_rcz} and  \ref{eq:spac_wigner}. This is evident in Figs \ref{gauss_block_asym_ph}, and \ref{wigner_asym_ph} respectively.

\begin{figure}[h]
\begin{center}
\includegraphics[scale=0.5]{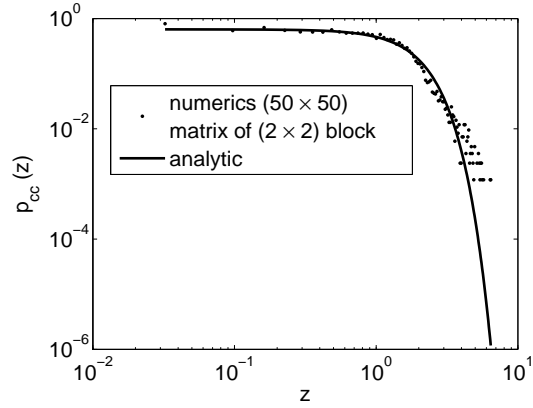}
\end{center}
\caption{Log-log plot of the distribution of spacing among complex-conjugate pairs. } \label{pccising}
\end{figure}

\begin{figure}[h]
\begin{center}
\includegraphics[scale=0.5]{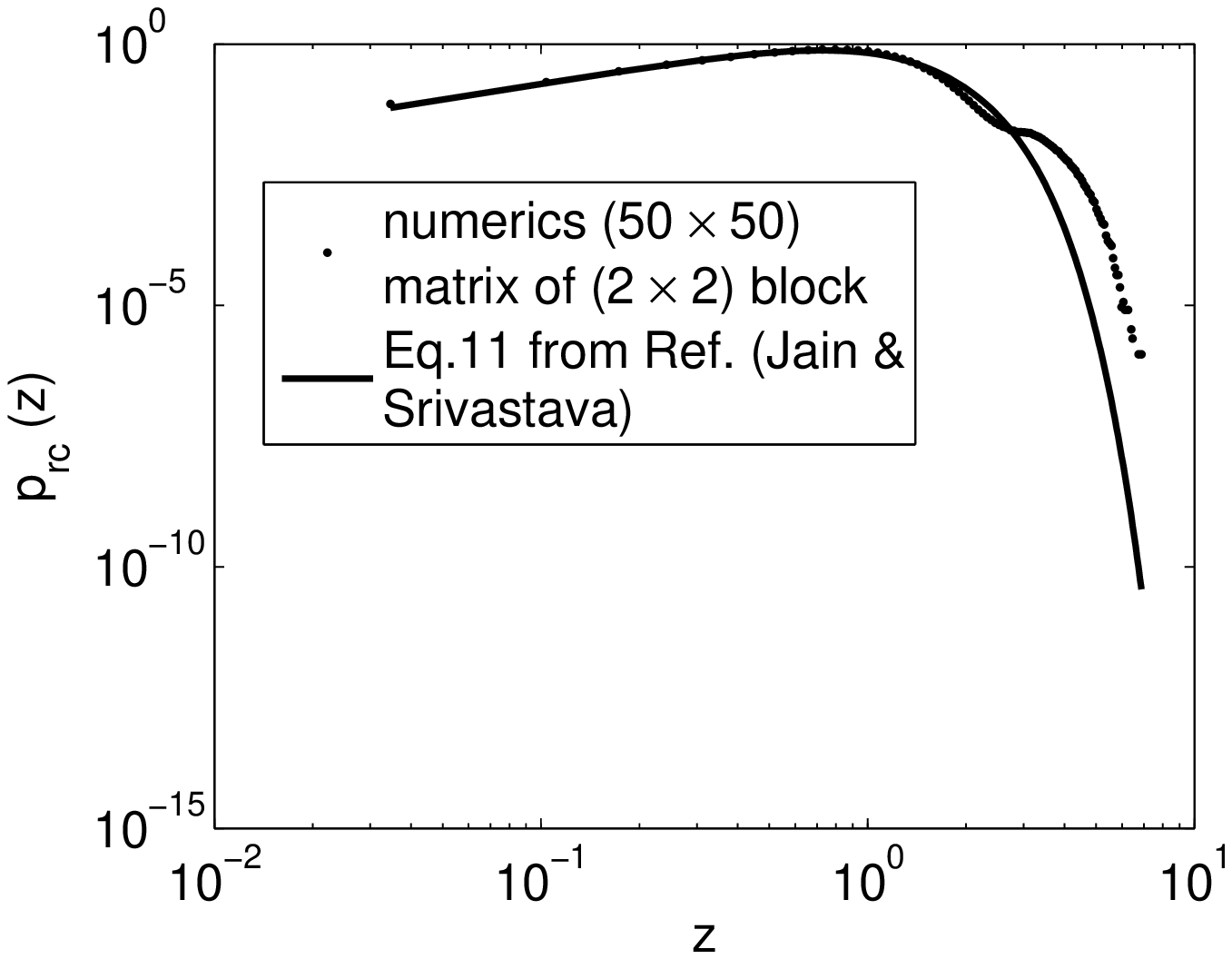}
\includegraphics[scale=0.5]{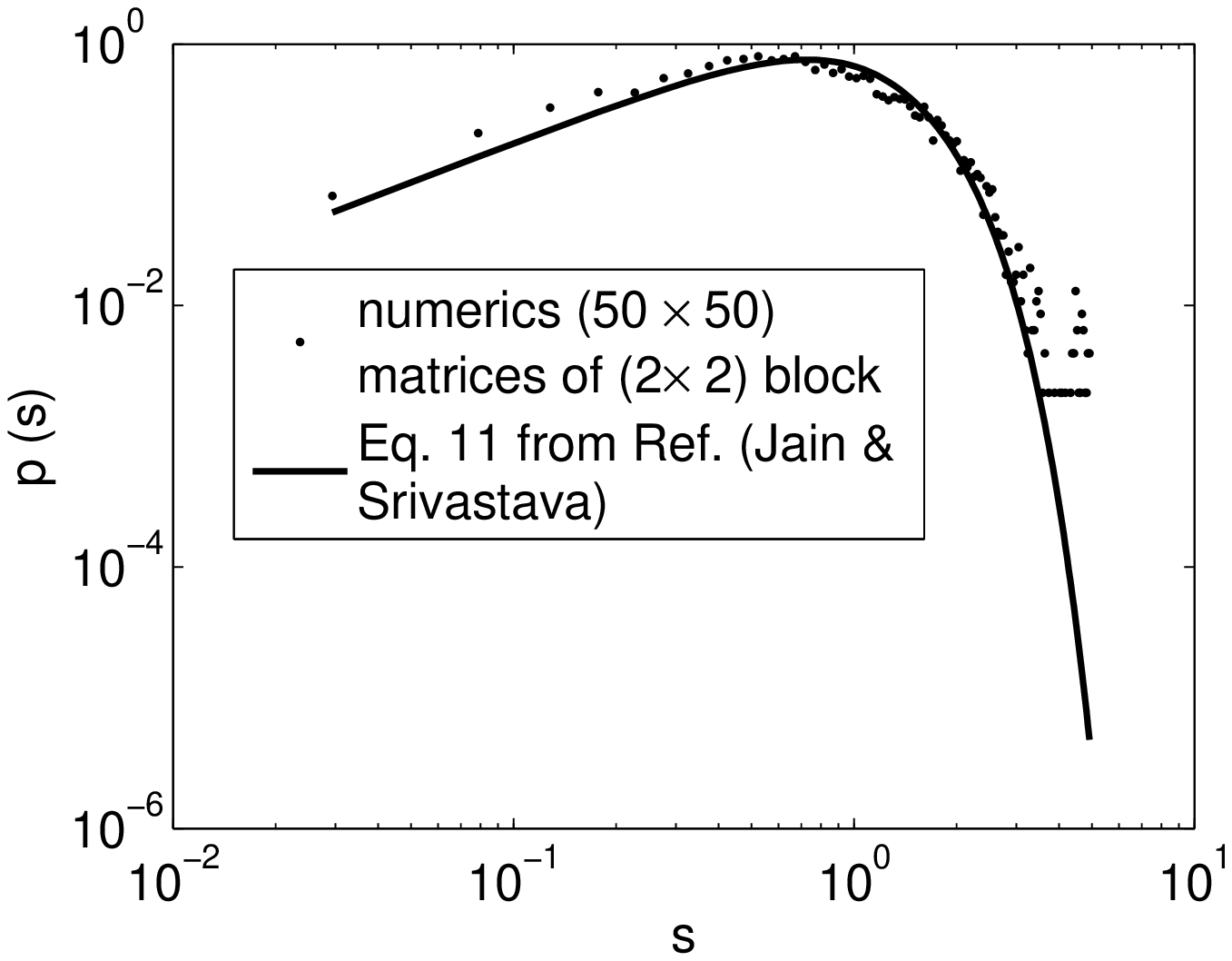}
\end{center}
\caption{(left) Log-log plot of the distribution of spacing among real and complex eigenvalues. (right) Log-log plot of the distribution of spacing among complex eigenvalues (Eq. \ref{eq:spac_rcz} is Eq. 11 from Ref. Jain \& Srivastava). }
\label{wignerising}
\end{figure}

The agreement is more intriguing than satisfactory as one would expect the block structure to show up in some form or the other in spectral distributions too. This led us to explore  the  cyclic blocks of $(2 \times 2)$ matrices having complex entries, numerically. In particular, keeping Ising model transition matrix in mind we focussed on cyclic matrices with the row $(A~~B~~B~~\ldots B^{\dagger})$ where 
\begin{eqnarray}
A = \left[\begin{array}{cc}a_1&ia_2\\-ia_2&a_1\end{array}\right],
\end{eqnarray}
\begin{eqnarray}
B = \left[\begin{array}{cc}-\frac{1}{2}&ib_1\\ib_2&-\frac{1}{2}\end{array}\right]
\end{eqnarray}
with real $a_1, a_2, b_1, b_2 $.  We have chosen the form of 
$A$ and $B$  from the structure appearing in the Ising model in two dimensions. The spacing distributions for different cases for random matrices are shown in Figs \ref{pccising}, \ref{wignerising}. 
The complex-conjugate pairs are spaced as shown in Fig. \ref{pccising} which is in reasonable agreement with Eq. \ref{eq:spac_cc}.  

The spacing distribution between real and complex eigenvalue shows a departure in the tail (Fig. \ref{wignerising}),
however we are yet to investigate the dependence of this deviation with the size of the matrix. Similar deviation of the numerical result from Wigner distribution  is seen in Fig. \ref{wignerising}.

This exploration throws a lot of interesting questions like obtaining the joint probability distribution function for the cyclic block models, and the ensuing spectral properties.

\section{Remarks}

Finally, we would like to point out a few significant areas we have not touched upon. There is a lot of research directed towards the physics beyond the standard model, particularly relevant to atomic and nuclear physics. In line with the theme of the Special Issue is atomic electric dipole moment, the discovery of which will be an instance of breakdown of parity and time-reversal symmetries. In certain nuclei with shapes that would support such a symmetry-scenario, it is expected that certain collective modes would enhance the Schiff moment \cite{vz}. The quadrupole and octupole collective modes could co-exist. This area would gain if many-body theory is systematically extended beyond mean-field and random phase approximations, to include ${\cal PT-}$ symmetry in the semiclassical descriptions. The description of nuclear dynamics requires a self-consistent coupling of Hamiltonian for nucleons and a Hamiltonian in the deformation space. This leads to rather non-trivial connection of chaos in single-particle 
motion, emergence of hydrodynamic  mode in deformation space giving rise to an effective inertia tensor \cite{jain2011}. 

Generally, there are much lesser known results on non-hermitian random matrix ensembles. The basic result is called the circular law, the analogue of the Wigner's semi-circular law for the density of eigenvalues.  This can be stated as spectral measure for $n\times n$ non-Hermitian iid matrices, whose elements are distributed in iid manner with fixed distribution (say Gaussian) with mean 0 and variance 1, converges to uniform measure in circle.  A figurative version will be eigenvalues if scaled properly of such a matrix will fill the unit circle uniformly. For Gaussian unitary non-Hermitian ensemble the joint probability density and correlation function were obtained by Ginibre \cite{Ginibre} while density of Ginibre orthogonal  ensemble was given by Sommers et al.\cite{sommers} and latter joint probability density function as well as the correlation function by Kanzieper and Akemann\cite{Kanzieper}. A nice review on random matrices close to Hermitian and unitary is by Fyodorov and 
Sommers \cite{fyodo}. In future, the assumption of joint independence of the matrix elements needs to be relaxed. Moreover, the non-Hermitian case presents the situation of spectral instability wherein a small perturbation in a large non-Hermitian matrix leads to large fluctuations. Similarly, there are a number of interesting problems that remain open.  For some new results on RMT for non-hermitian systems, we refer the reader to the work by Tao \cite{terry}.      

 The results on random ${\cal PT}$-symmetric systems are expected to be relevant for a wide variety of physical situations occurring in anyon physics \cite{halperin}, $\nu  $-parametrized quantum chromodynamics, fractional quantum hall systems \cite{jkjain}, etc. The connection of RMT with the momentum distribution functions was established for the Maxwell-Boltzmann, Fermi-Dirac, and Bose-Einstein cases \cite{mark,alonso}. As pointed out in \cite{alonso,jalonso}, the momentum distribution function for the anyon gas requires the development of general pseudo-Hermitian RMT. Some current development related with ${\cal PT}$-symmetric deformations of integrable models is reviewed by Fring\cite{Fring} while another interesting article on supersymmetric many-particle quantum systems with inverse-square interactions are now available\cite{Ghosh}.

Non-hermitian circular billiard is shown to possess energy levels that show some level repulsion, alongwith the eigenfunctions which are directional \cite{saket}. Further work on these lines will be relevant for the design of micro-disk lasers.

\end{document}